\documentclass[aps,prd,preprint,showpacs,showkeys,preprintnumbers,amsmath,amssymb]{revtex4-1}
\usepackage{amsmath}
\usepackage{amssymb}

\begin{document}

\title{On the Quantization of the Charge-Mass Ratio}
\author{S. C. Ulhoa }
\email{sc.ulhoa@gmail.com} \affiliation{Instituto de F\'{i}sica,
Universidade de Bras\'{i}lia, 70910-900, Bras\'{i}lia, DF,
Brazil.}

\date{\today}

\begin{abstract}
The paper deals with the problem of describing fundamental particles. The Einstein-Rosen approach was revisited to explain que charge-mass ratio quantization. Such a result is obtained once a quantization prescription is applied to the expression of gravitational energy defined in the realm of teleparallel gravity.
\end{abstract}



\maketitle
\section{Introduction}
\noindent

In 1935 A. Einstein and N. Rosen proposed a theory to describe fundamental particles in terms of geometry of space-time~\cite{PhysRev.48.73}. Thus particles wold be a bridge between two sheets in space-time which is known as the Einstein-Rosen bridge. Such a structure was later associated to wormholes. In this context the field equations slightly different from those of general relativity was used to describe particles, it is due to a superposition aspect of nature. Macroscopic masses are described by such equations and a body with mass is composed by particles, thus it is natural to describe particles by the same set of equations. The difference is a minus sign to avoid singularities. Hence a fundamental charged particle was described by a regular version of Reissner-Nordstrom metric. The authors of reference \cite{PhysRev.48.73} didn't seem to believe that solutions with singularities could be associated to elementary particles since the presence of such a feature would bring too much arbitrariness to Physics. They also made an interesting criticism of this approach: charge and mass are independent quantities that come from constants in the integration. There is no evidence of such independence since there is no observed massless charged particle. This idea was abandoned later because it was difficult to make predictions about the internal structure of particles or a system of particles and more difficult to measure such predictions. This approach was unable to explain fundamental relations between charge and mass such as the quantization of the charge-mass ratio.

P. Dirac~\cite{PhysRev.74.817} showed that the quantization of the charge-mass ratio was a direct consequence of the existence of a magnetic monopole which has never been observed. However this approach tells nothing about an internal structure of a charged particle. Thus fundamental particles are described in Physics by mechanical points. It is a strange viewpoint because a structure with mass cannot be a point. Hence the conclusion is clear: this viewpoint is only an approximation. The problem on how to proper describe fundamental particles remains as one of the most interesting challenges in Physics.

I believe that a charged fundamental particle should have an internal structure which is described by a geometry coupled to electromagnetic field, hence it is promising to use Einstein-Rosen approach conjugated to a quantization process. In this way the predictions of the theory will lay on the observables which are precisely mass and charge. I show in this article that a quantum version of Einstein-Rosen idea leads naturally to a quantized charge-mass ratio. Thus the metric of the fundamental particle would yields a quantity analogous to the wave function of quantum mechanics once a quantization prescription is introduced. On the other hand in the metric formulation of general relativity it is not clear what quantity should be quantized since the natural candidate, the energy, is not well defined in this context. The same is not true for teleparallel gravity in which is possible to define a reliable expression for gravitational energy. Thus such a quantity is used in this article to explain the quantization of the charge-mass ratio.

This article is divided as follows. In section \ref{tel} it is introduced the main ideas of teleparallel gravity. Then in section \ref{CM} the quantization of charge-mass ratio is discussed. Such a result comes from a geometric theory of particle conjugated to a quantization process. Finally the concluding remarks are presented in the last section.  In addition we adopt units where
$G=c=1$, unless otherwise stated.

\section{Teleparallel
Equivalent to General Gelativity (TEGR)}\label{tel} \noindent

In this section we present some basic ideas of Teleparallel gravity which is an alternative theory of gravitation and dynamically equivalent to general relativity. In such a theory the tetrad field plays the role of the dynamical variable instead of the usual metric tensor. It was introduced by A. Einstein in the 1930's as a first step towards a unified field theory~\cite{Einstein}. The tetrad field $e^a\,_\mu$ is endowed with two symmetries: Lorentz symmetry which is represented by Latin indices and diffeomorphism symmetry which is represented by Greek indices. Hence $\mu=0,i$ and $a=(0),(i)$. In this sense the tetrad field projects tensors under coordinate transformations into tensors under Lorentz transformations. The metric tensor components are related to the tetrad field by the relation $g_{\mu\nu}=e^a\,_\mu e_{a\nu}$ which means that general relativity can be formulated in terms of tetrads. However there is an apparent contradiction here the metric is symmetric which leaves 10 independent components, on the other hand the tetrad field has 16 components. It should be noted that the component $e_{(0)}\,^\mu$ is always tangent to the world line of the observer, thus this is interpreted as the field velocity of the observer. Therefore this extra components in the tetrad field are linked to the freedom in the choice of the reference frame. In this sense the tetrad field is adapted to a specific observer.

General relativity in its metric formulation is established into a Riemannian geometry in which the curvature is constructed out of a torsion-free connection, the Christoffel symbols ${}^0\Gamma_{\mu \lambda\nu}$. It is possible to relate such a geometry to the Weitzenb\"ock geometry in which there is a non-vanishing torsion tensor that is constructed out of a curvature-free connection $\Gamma_{\mu \lambda\nu}$, also known as Cartan connection. This connection is explicitly given by $\Gamma_{\mu \lambda\nu}=e_{a\mu}\partial_{\lambda} e^{a}\,_{\nu}$, hence it defines a torsion as

\begin{equation}
T^{a}\,_{\lambda\nu}=\partial_{\lambda} e^{a}\,_{\nu}-\partial_{\nu}
e^{a}\,_{\lambda}\,. \label{3}
\end{equation}

The Cartan connection satisfies the following identity
\begin{equation}
\Gamma_{\mu \lambda\nu}= {}^0\Gamma_{\mu \lambda\nu}+ K_{\mu
\lambda\nu}\,, \label{2}
\end{equation}
where

\begin{eqnarray}
K_{\mu\lambda\nu}&=&\frac{1}{2}(T_{\lambda\mu\nu}+T_{\nu\lambda\mu}+T_{\mu\lambda\nu})\label{3.5}
\end{eqnarray}
is the contortion tensor. Thus the next step is associate the curvature scalar $R({}^0\Gamma)$ in Riemannian geometry to some expression in the Weitzenb\"ock geometry which is accomplished with the use of identity (\ref{2}). Then we recall that the curvature scalar calculated with Cartan connection vanishes identically, this yields

\begin{equation}
eR({}^0\Gamma)\equiv -e({1\over 4}T^{abc}T_{abc}+{1\over
2}T^{abc}T_{bac}-T^aT_a)+2\partial_\mu(eT^\mu)\,,\label{5}
\end{equation}
where $e$ is the determinant of the tetrad field and $T_a=T^b\,_{ba}$
$\left(T_{abc}=e_b\,^\mu e_c\,^\nu T_{a\mu\nu}\right)$.
Therefore a gravitational theory equivalent to general relativity can be established by means the following Lagrangian density
\begin{equation}
\mathfrak{L}= -k e({1\over 4}T^{abc}T_{abc}+{1\over
2}T^{abc}T_{bac}- T^aT_a) -\mathfrak{L}_M \label{lag}\,,
\end{equation}
where $k=1/16\pi$ and $\mathfrak{L}_M$ stands for the Lagrangian
density of matter fields. This is precisely the case of teleparallel gravity. It should be noted that a total divergence in the Lagrangian density does not alter the field equations. It is interesting  to rewrite (\ref{lag}) as

\begin{equation}
\mathfrak{L}\equiv -ke\Sigma^{abc}T_{abc} -\mathfrak{L}_M\,,
\label{5}
\end{equation}
where

\begin{equation}
\Sigma^{abc}={1\over 4} (T^{abc}+T^{bac}-T^{cab}) +{1\over 2}(
\eta^{ac}T^b-\eta^{ab}T^c)\,. \label{6}
\end{equation}
Thus we can perform a variation with respect to the tetrad field which yields the field equations. They read

\begin{equation}
e_{a\lambda}e_{b\mu}\partial_\nu (e\Sigma^{b\lambda \nu} )- e
(\Sigma^{b\nu}\,_aT_{b\nu\mu}- {1\over 4}e_{a\mu}T_{bcd}\Sigma^{bcd}
)={1\over {4k}}eT_{a\mu}\,, \label{7}
\end{equation}
where $\delta \mathfrak{L}_M / \delta e^{a\mu}=eT_{a\mu}$. Such equations are equivalent to Einstein equations. In this sense every known solution in general relativity will also be a solution in teleparallel gravity. However both theories do not share all features. For instance the definition of gravitational energy remains problematic, on the other hand in teleparallel gravity there is a reliable definition of such a quantity.

In order to define energy let us rewrite equation (\ref{7}) as
\begin{equation}
\partial_\nu(e\Sigma^{a\lambda\nu})={1\over {4k}}
e\, e^a\,_\mu( t^{\lambda \mu} + T^{\lambda \mu})\;, \label{8}
\end{equation}
where $T^{\lambda\mu}=e_a\,^{\lambda}T^{a\mu}$ and

\begin{equation}
t^{\lambda \mu}=k(4\Sigma^{bc\lambda}T_{bc}\,^\mu- g^{\lambda
\mu}\Sigma^{bcd}T_{bcd})\,. \label{9}
\end{equation}
Due to the antisymmetry $\Sigma^{a\mu\nu}=-\Sigma^{a\nu\mu}$, it follows that

\begin{equation}
\partial_\lambda
\left[e\, e^a\,_\mu( t^{\lambda \mu} + T^{\lambda \mu})\right]=0\,.
\label{10}
\end{equation}
This is a local conservation equation. As a consequence we get

\begin{equation}
{d\over {dt}} \int_V d^3x\,e\,e^a\,_\mu (t^{0\mu} +T^{0\mu})
=-\oint_S dS_j\, \left[e\,e^a\,_\mu (t^{j\mu} +T^{j\mu})\right]\,.
\label{11}
\end{equation}
Therefore we identify $t^{\lambda\mu}$ as the gravitational
energy-momentum tensor~\cite{PhysRevLett.84.4533,maluf2}.

Thus the energy-momentum vector is defined as~\cite{Maluf:2002zc}

\begin{equation}
P^a=\int_V d^3x\,e\,e^a\,_\mu (t^{0\mu} +T^{0\mu})\,, \label{12}
\end{equation}
where $V$ is a volume of the three-dimensional space. Some features of such expression should be noted. Firstly it is independent on the coordinate system which is desirable for each definition of energy. Secondly it is dependent on the choice of the reference frame since it is a vector under Lorentz transformations. This dependence appears in special relativity as well, there the energy vary from $mc^2$ to $\gamma mc^2$ depending on the reference frame. Therefore a definition of gravitational energy should take this feature into account.

\section{Quatization of Charge-Mass Ratio}\label{CM}
\noindent
In this section we intent to show how to obtain the quantized charge-mass ratio. In order to accomplish such a goal we recover Einstein's idea about the role of curvature in the description of fundamental particles~\cite{PhysRev.48.73}. We imagine that a spinless charged particle could be described by the Reissner-Nordstrom metric
\begin{equation}
ds^2=-f(r)dt^2+f(r)dr^2+r^2d\theta^2+r^2\sin^2\theta d\phi^2\,,\label{metrica}
\end{equation}
with $f(r)=1-\frac{2M}{r}+\frac{Q^2}{r^2}$. This description was followed by M. O. Katanaev who demonstrated that a particle with mass $m$ can be described by Schwarzschild metric in isotropic coordinates~\cite{Katanaev:2012wc}. Thus the internal structure of such a particle will lay in its geometry. Let us choose a reference frame adapted to an observer at rest which is realized by the following tetrad field
\begin{equation}
e^a\,_{\mu}=\left(
  \begin{array}{cccc}
    \sqrt{-g_{00}}&0&0&0 \\
    0&\sqrt{g_{11}}
\,\sin\theta \cos\phi& \sqrt{g_{22}} \cos\theta \cos\phi & -\sqrt{g_{33}}\sin\phi \\
    0& \sqrt{g_{11}}\, \sin\theta \sin\phi&
\sqrt{g_{22}} \cos\theta \sin\phi &  \sqrt{g_{33}} \cos\phi \\
    0&
\sqrt{g_{11}}\, \cos\theta & -\sqrt{g_{22}}\sin\theta&0 \\
  \end{array}
\right)\,,\label{tetrada}
\end{equation}
the energy density associated to this tetrad is
\begin{eqnarray}
4e\Sigma^{(0)01}&=&2\left(\sqrt{g_{33}}+\sqrt{g_{22}}\sin\theta\right)-\frac{1}{\sqrt{g_{11}}}\left[\sqrt{\frac{g_{33}}{g_{22}}}\left(\frac{\partial g_{22}}{\partial r}\right)+\sqrt{\frac{g_{22}}{g_{33}}}\left(\frac{\partial g_{33}}{\partial r}\right)\right]\,,
\end{eqnarray}
which specializes into
\begin{equation*}
4e\Sigma^{(0)01}=4r\sin\theta\left[1-\left(1-\frac{2M}{r}+\frac{Q^2}{r^2}\right)^{1/2}\right]\,,
\end{equation*}
once we substitute the above metric. It should be noticed that the total energy of the space-time is given by

\begin{eqnarray}
P^{(0)}=E&=&\lim_{r\rightarrow\infty}\int_0^{2\pi}\int_0^{\pi}4e\Sigma^{(0)01}d\theta d\phi\nonumber\\
&=&M\,.
\end{eqnarray}

The next step is to obtain a quantized description of the system. This process of mapping a classical system into a quantum one is called quantization procedure. The essence of any quantization is to introduce non-commutative variables which replace the commutative ones. For instance the Schr\"odinger equation is obtained as a result of a quantization procedure in the classical phase space. Thus the variables are replaced by operators following a certain prescription which in the case of Schr\"odinger equation is given by $x^i\rightarrow \hat{x^i}=x^i$ and $p_i\rightarrow \hat{p_i}=-i\frac{\partial}{\partial x^i}$. However there is no consensus about how to establish such a quantization procedure. Probably  the most accepted way is the so called Weyl quantization prescription~\cite{weyl} which we have recently used to obtain a quantized spectrum of the mass of Schwarzschild black hole~\cite{Ad.H.Phys.2014}. As a matter of fact the Weyl prescription yields the correct form of the variables in Schr\"odinger equation. Inspired by this prescription we propose:
$\sin\theta\rightarrow\hat{\sin\theta}=\alpha$ and $r\rightarrow\hat{r}=\beta\frac{\partial}{\partial\alpha}$, where $\beta$ is a constant analogous to the Planck's constant. It is interesting to note that the quantization procedure applied to a geometric structure induces to a space-time with non-commutative variables. The commutator between $\rightarrow\hat{r}$ and $\hat{\sin\theta}$ is $\beta$, thus $\beta<<1$ once this non-commutative property is not observed in everyday life. As a consequence $4e\Sigma^{(0)01}
\rightarrow\hat{\mathcal{H}}$. In order to avoid problems with the operator ordering, let us symmetrize the Hamiltonian density. Hence

\begin{equation}
\hat{\mathcal{H}}=2\hat{r}\hat{\sin\theta}+2\hat{\sin\theta}\hat{r}-2\hat{\sin\theta}\left(\hat{r}^2-2M\hat{r}+Q^2\right)^{1/2}-2\left(\hat{r}^2-2M\hat{r}+Q^2\right)^{1/2}\hat{\sin\theta}\,.
\end{equation}

If we use $\beta<<1$ and look for an equation as $\hat{\mathcal{H}}\psi=\epsilon\psi$, then we find

\begin{equation}
-\frac{4\beta^2\alpha}{2Q}\frac{\partial^2\psi}{\partial\alpha^2}+4\beta\left[\alpha\left(1+\frac{M}{Q}\right)
-\frac{\beta}{2Q}\right]\frac{\partial\psi}{\partial\alpha}+\left[2\beta\left(1+\frac{M}{Q}\right)-\epsilon-4Q\alpha\right]\psi=0\,,
\end{equation}
the solution of this equation is given by
$$\psi=U(\alpha)\psi_0\exp{\left[\left(\frac{\alpha}{\beta}\right)\left(Q+M-\sqrt{M^2+2QM-Q^2}\right)\right]}\,,$$
where $U(\alpha)=F(a,1,x)$, with $x=\left(\frac{2\sqrt{M^2+2QM-Q^2}}{\beta}\right)\alpha$ and $$a=-\frac{1}{4}\left(\frac{\epsilon Q-2\beta \sqrt{M^2+2QM-Q^2}}{\beta \sqrt{M^2+2QM-Q^2}}\right)\,.$$ It turns out that $F(a,1,x)$ is the Kummer function that obeys the following differential equation

$$xF^{''}+(b-x)F^{'}-aF=0\,,$$
with $F^{'}\equiv \frac{\partial F}{\partial x}$ thus $F\equiv F(a,b,x)$.

If $a=n$ with $n$ an integer then $F(a,1,x)$ will be finite, such a condition implies that

\begin{equation}
\epsilon=2\beta(2n+1)\sqrt{\frac{M^2}{Q^2}+2\frac{M}{Q}-1}\,.\label{epsilon}
\end{equation}
The total energy is
$$E=\int\psi^\dag\hat{\mathcal{H}}\psi d^2x=\epsilon\,,$$
this quantity is an observable which is $E=M$. Thus using this with expression (\ref{epsilon}), the charge-mass ratio is given by

\begin{equation}
\frac{Q}{M}=\left[\frac{2n+1}{-(2n+1)\pm\sqrt{2(2n+1)^2+m^2}}\right]\,,
\end{equation}
where $m^2=\frac{M^2}{4\beta^2}$. Therefore the charge-mass ratio is quantized as a consequence of non-commutative variables in a Reissner-Nordstrom geometry that describes a spinless charged particle. If $m>>n$ then the above ratio simplifies to $Q/M=\pm (2n+1)/m$.

\section{Conclusion}

In this article the Einstein-Rosen approach to describe fundamental particles has been revisited. Particles are viewed as a space-time geometry established by Einstein equation. This viewpoint is understood as the very nature of gravitational field which is a macroscopic manifestation of such a geometric particle theory. In particular charged particles is believed to be described by Reissner-Nordstrom metric. As a consequence a quantization procedure applied to this system leads to a quantization of the charge-mass ratio. In such a process a constant analogous to Planck constant is introduced and for the condition $m>>n$ it seems to be the very electron's charge which is the fundamental charge or in SI units $\beta=\left(\sqrt{\frac{G}{4\pi\epsilon_0 c^4}}\right)e$. Thus it is interesting to analyze what the theory predicts for the electron itself. In my opinion the spin should have a close relation to torsion.


%

\end{document}